\documentclass[runningheads]{llncs}
\usepackage{graphicx}
\begin{document}
\title{Scaling Cross-Domain Content-Based Image Retrieval for E-commerce Snap and Search Application}
%
%
\author{Isaac Chung\inst{1} \and
Minh Tran\inst{2} \and
Eran Nussinovitch\inst{2}}
\authorrunning{I. Chung et al.}
%
\institute{Clarifai, Tallinn, Estonia \and
Clarifai, New York, USA\\
\email{first.last@clarifai.com}\\
}
\maketitle              
\begin{abstract}
In this industry talk at ECIR 2022, we illustrate how we approach the main challenges from large scale cross-domain content-based image retrieval using a cascade method and a combination of our visual search and classification capabilities. 
Specifically, we present a system that is able to handle the scale of the data for e-commerce usage and the cross-domain nature of the query and gallery image pools. 
We showcase the approach applied in real-world e-commerce snap and search use case and its impact on ranking and latency performance. 

\end{abstract}
\section{Introduction}
\begin{figure} 
    \centering
    \includegraphics[scale=0.06]{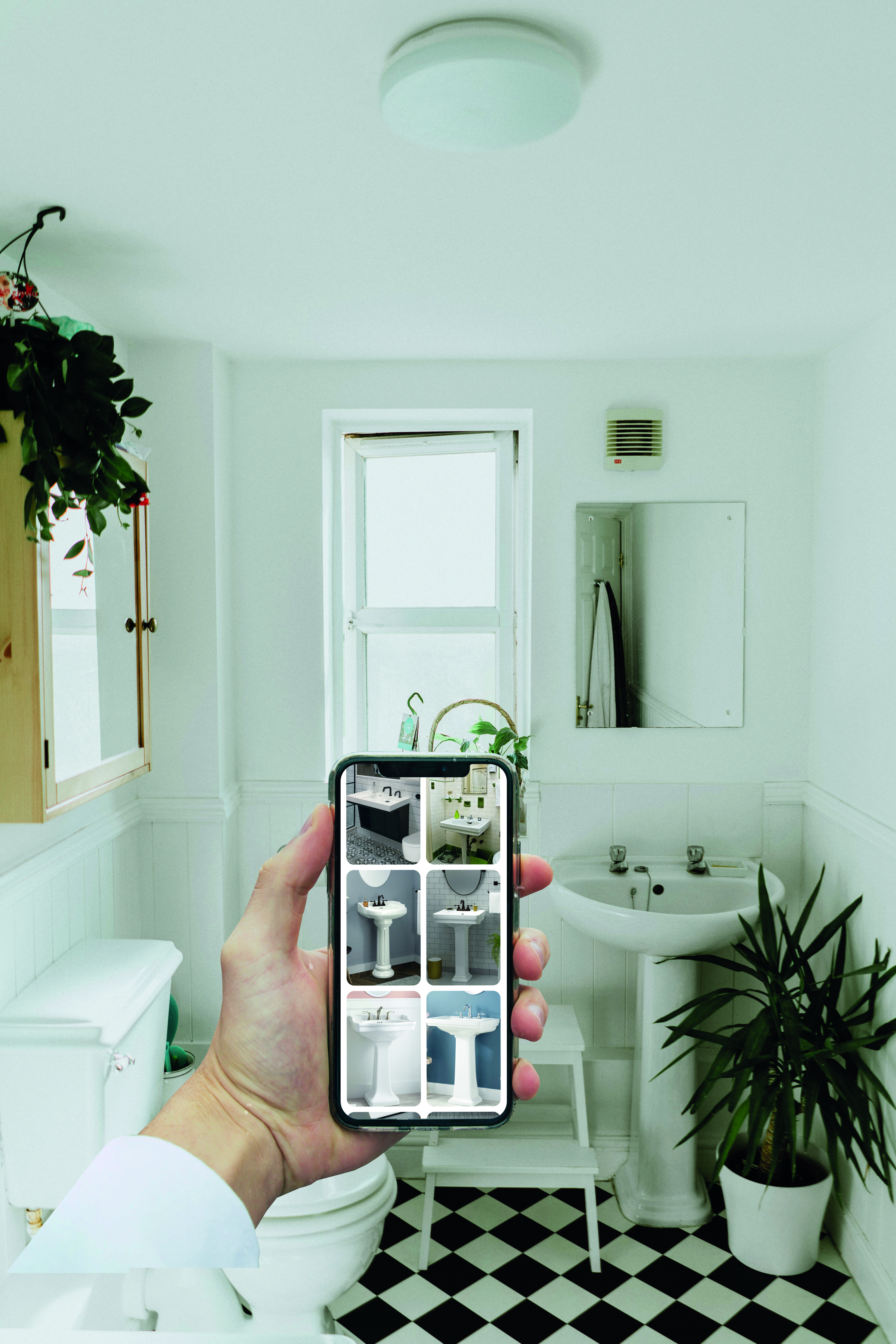}
    \caption{Snap and Search for eCommerce}
    \label{fig:vs1}
\end{figure}

E-commerce companies are increasingly adopting Computer Vision based technologies to offer a more appealing shopping experience and reduce shopper bounce rate \cite{etsy,zhang2018visual,shankar2017deep,Stanley2020SIRSI}. Visual search recommendations in their mobile and web applications, illustrated in Figure \ref{fig:vs1}, improve customer experience by reducing the time and effort needed for product searches. These efforts aim to grow the number of conversion opportunities, which would lead to increased revenue.

\section{Methods}
\begin{figure} 
    \centering
    \includegraphics[scale=0.7]{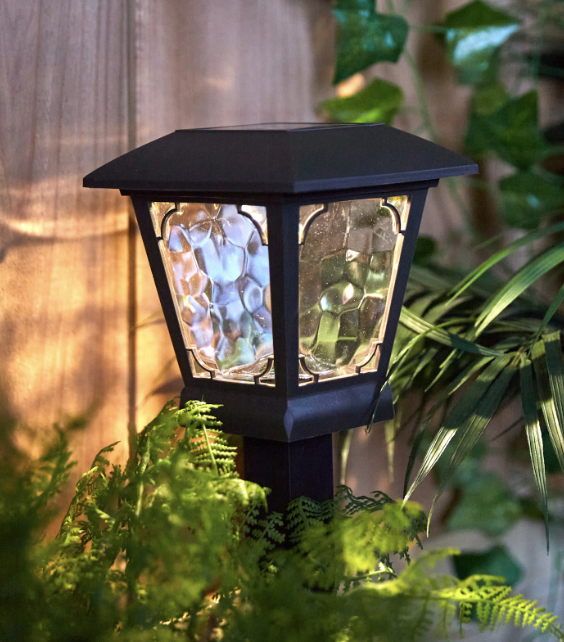}
    \includegraphics[scale=0.7]{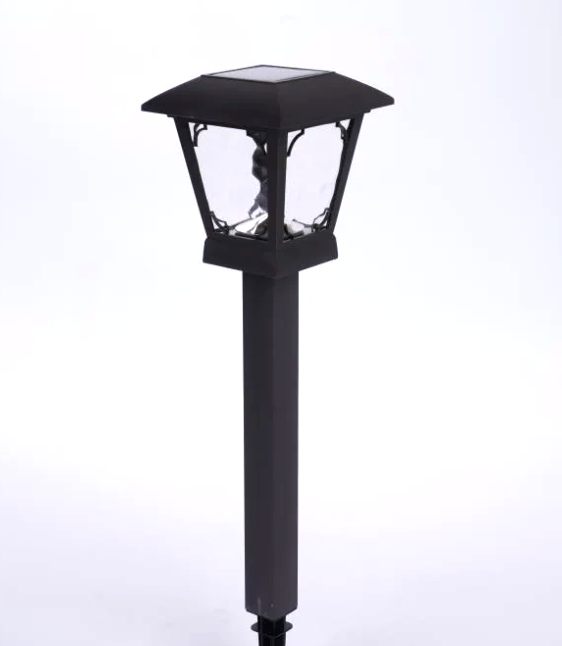}
    \caption{Example Query (top) and Catalogue (bottom) Images.}
    \label{fig:vs2}
\end{figure}

We achieve improvements using a combination of visual search and classification capabilities, working in tandem with a large e-commerce client. This allows users to use their own photos to search for products via the client’s mobile app. This content-based image retrieval task presents a couple of problems. First, the gallery set contains 2M+ images which span 3000+ products and 30+ top level categories (TLC). These categories were not designed with visual search in mind, but rather with traditional shopping, e.g. socks can show up under mens or womens top level categories, shelves could be in a bedroom or a garage, faucets can show up in both bathroom and kitchen. With a product catalogue this vast, seemingly similar looking items with functionally different uses can be confused with one another without the appropriate, overarching context. Applying brute force search algorithms at this scale fails if the dimensions in the latent space are not carefully tuned. Second, the gallery set images come from a different domain than the user generated (query) images, as shown in Figure \ref{fig:vs2}: Those to be retrieved are often studio-taken photos with a white background, or photos of products staged in their intended environment, while the users' query images are smartphone photos, which can have noisy backgrounds as well as large variations in orientation and lighting. This introduces complexities in comparing between image representations.

Our approach addresses such problems by introducing an intermediate stage between the feature extractor and the approximate nearest neighbors (ANN) algorithm in the image retrieval system. This stage leverages the granular product hierarchy to perform a cascade-style search. The gallery set is partitioned by TLC. A multi-class classifier, which is a convolutional neural network (CNN) trained on images in the same domain as the query data with labels as their respective TLCs, takes in query images as input and outputs TLC as predictions. Search is then performed in the gallery set partition of the top prediction. With this, we are able to improve image retrieval metrics by an average of 69.7\% compared to our baseline method, while limiting the overall latency increase to only 13\%.

\section{Company Portrait and Speakers}
\paragraph{Company} Clarifai is a leading provider of artificial intelligence for unstructured data. We help organizations transform their images, video, text, and audio data into actionable insights at scale by providing an AI ecosystem for developers, data scientists, and no-code operators. Clarifai supports the full AI development lifecycle; including dataset preparation, model training and deployment. Founded in 2013, Clarifai continues to grow with employees remotely based throughout the United States and Estonia. 

\paragraph{Speaker} Isaac Chung joined Clarifai in the Fall of 2020 as a Machine Learning Engineer. He focuses on technical deliveries in commercial projects, especially on visual search applications. He holds a B.A.Sc in Engineering Science (Aerospace) from the University of Toronto. Before pursuing his M.A.Sc in Machine Learning also from the University of Toronto, he spent 2 years in the aerospace industry in Safran Landing Systems.

%
%
%
\bibliographystyle{splncs04}
%

\bibliography{references}





\end{document}